\documentclass[a4paper,fleqn,usenatbib]{mnras}

\usepackage[T1]{fontenc}
\usepackage{ae,aecompl}

\usepackage{graphicx}	
\usepackage{amsmath}	
\usepackage{amssymb}	

\newcommand{\kms}{\,km\,s$^{-1}$}

\title[H\,{\normalsize \textbf{I}} absorption structure in a low-redshift galaxy]{Parsec-scale H\,{\normalsize \textbf{I}} absorption structure in a low-redshift galaxy seen against a Compact Symmetric Object}

\author[A. D. Biggs et al.]{
  A.~D.~Biggs,$^{1}$\thanks{E--mail: abiggs@eso.org}
  M.~A.~Zwaan,$^{1}$
  E. Hatziminaoglou,$^{1}$
  C. P\'{e}roux$^{2}$
  and J.~Liske$^{1,3}$
  \\
  $^{1}$European Southern Observatory, Karl Schwarzschild Stra{\ss}e 2, D-85748 Garching, Germany \\
  $^{2}$Aix Marseille Universit\'{e}, CNRS, LAM (Laboratoire d'Astrophysique de Marseille) UMR 7326, 13388, Marseille, France \\
  $^{3}$Hamburger Sternwarte, Universit{\"a}t Hamburg, Gojenbergsweg 112, 21029 Hamburg, Germany
}

\date{Accepted XXX. Received YYY; in original form ZZZ}

\pubyear{2016}

\begin{document}
\label{firstpage}
\pagerange{\pageref{firstpage}--\pageref{lastpage}}
\maketitle

\begin{abstract}
  We present global VLBI observations of the 21-cm transition of atomic hydrogen seen in absorption against the radio source J0855+5751. The foreground absorber (SDSS~J085519.05+575140.7) is a dwarf galaxy at $z = 0.026$. As the background source is heavily resolved by VLBI, the data allow us to map the properties of the foreground \ion{H}{i} gas with a spatial resolution of 2~pc. The absorbing gas corresponds to a single coherent structure with an extent $>$35~pc, but we also detect significant and coherent variations, including a change in the \ion{H}{i} optical depth by a factor of five across a distance of $\la 6$~pc. The large size of the structure provides support for the \citeauthor{heiles03} model of the ISM, as well as its applicability to external galaxies. The large variations in \ion{H}{i} optical depth also suggest that caution should be applied when interpreting $T_S$ measurements from radio-detected DLAs. In addition, the distorted appearance of the background radio source is indicative of a strong jet-cloud interaction in its host galaxy. We have measured its redshift ($z = 0.54186$) using optical spectroscopy on the William Herschel Telescope and this confirms that J0855+5751 is a FRII radio source with a physical extent of $<$1~kpc and supports the previous identification of this source as a Compact Symmetric Object. These sources often show absorption associated with the host galaxy and we suggest that both \ion{H}{i} and OH should be searched for in J0855+5751.
\end{abstract}




\section{Introduction}
\label{introduction}

Much of what we know about the gas composition of galaxies has been made possible through observations of atomic hydrogen in absorption. In contrast to emission lines, the minimum optical depth that can be detected is set by the brightness of the background source and thus detections are almost independent of redshift. Detection of Ly-$\alpha$ absorption at optical and ultraviolet wavelengths has been very successful at providing information on the physical state of a wide variety of media, from the low-density intergalactic medium to high-redshift galaxies \citep*[e.g.][]{bechtold03,wolfe05,noterdaeme12}. A complementary probe is the \ion{H}{i} hyperfine 21-cm line, this having the additional advantage that it does not saturate and is unaffected by dust obscuration. The 21-cm line enables a determination of the kinematics and gas distribution in the intervening absorbers, which should then allow these systems to be linked to their $z=0$ galaxy counterparts seen in 21-cm emission \citep{zwaan05a}.

One application of 21-cm absorption line studies that has been poorly exploited is the study of small-scale structure in the interstellar medium (ISM) of external galaxies. Observations of background structures on the order of tens of mas will probe the ISM of foreground galaxies at a distance of $\sim$100~Mpc on scales of tens of pc. This is interesting because it bridges the gap between the maximum of a few pc that can be studied in the ISM of the Milky Way in absorption against background radio sources \citep[e.g.][]{roy10}, and the more than 100-pc scales that are probed by high resolution 21-cm emission line maps of nearby galaxies \citep*[e.g.][]{elmegreen01,zhang12}.

Neutral hydrogen opacity fluctuations on parsec scales provide us with information on the processes that regulate star formation. The structure and turbulence of the neutral medium eventually determine the size distribution of molecular clouds and affects the shape of the stellar initial mass function \citep{padoan02}. Therefore, measuring the small-scale structure of the ISM is essential for our understanding of how cold gas is converted into stars over cosmic time. \citet*{mckee77} modelled the ISM as a collection of small (0.4-10~pc) isotropic clouds, each containing a central cold core and a warmer envelope -- these two components constitute the cold and warm neutral media (CNM and WNM). More recently, an extensive analysis of Galactic 21-cm absorption lines led \citet{heiles03} to conclude that the ISM can be modelled by ``blobby sheets'' i.e.\ the CNM consists of sheet-like structures with blobs (cloudlets) embedded within. 

As an example of the study of an individual background source, \citet{srianand13} have presented a case of a quasar sight line piercing the gas disk of a galaxy at a redshift of $z=0.08$. Very Long Baseline Array (VLBA) observations resulted in 21-cm absorption spectra toward three components separated by $\sim$10 and 90~pc at the distance of the galaxy. The measured optical depths toward the two components differ by up to a factor of 10 for the narrowest components and much less for the broader component. This is interpreted as being due to small ($<10$~pc) dense clouds embedded in a diffuse neutral medium on scales of several tens of parsecs, in support of the \citeauthor{heiles03} model.

\begin{figure*}
  \begin{center}
    \includegraphics[scale=0.15]{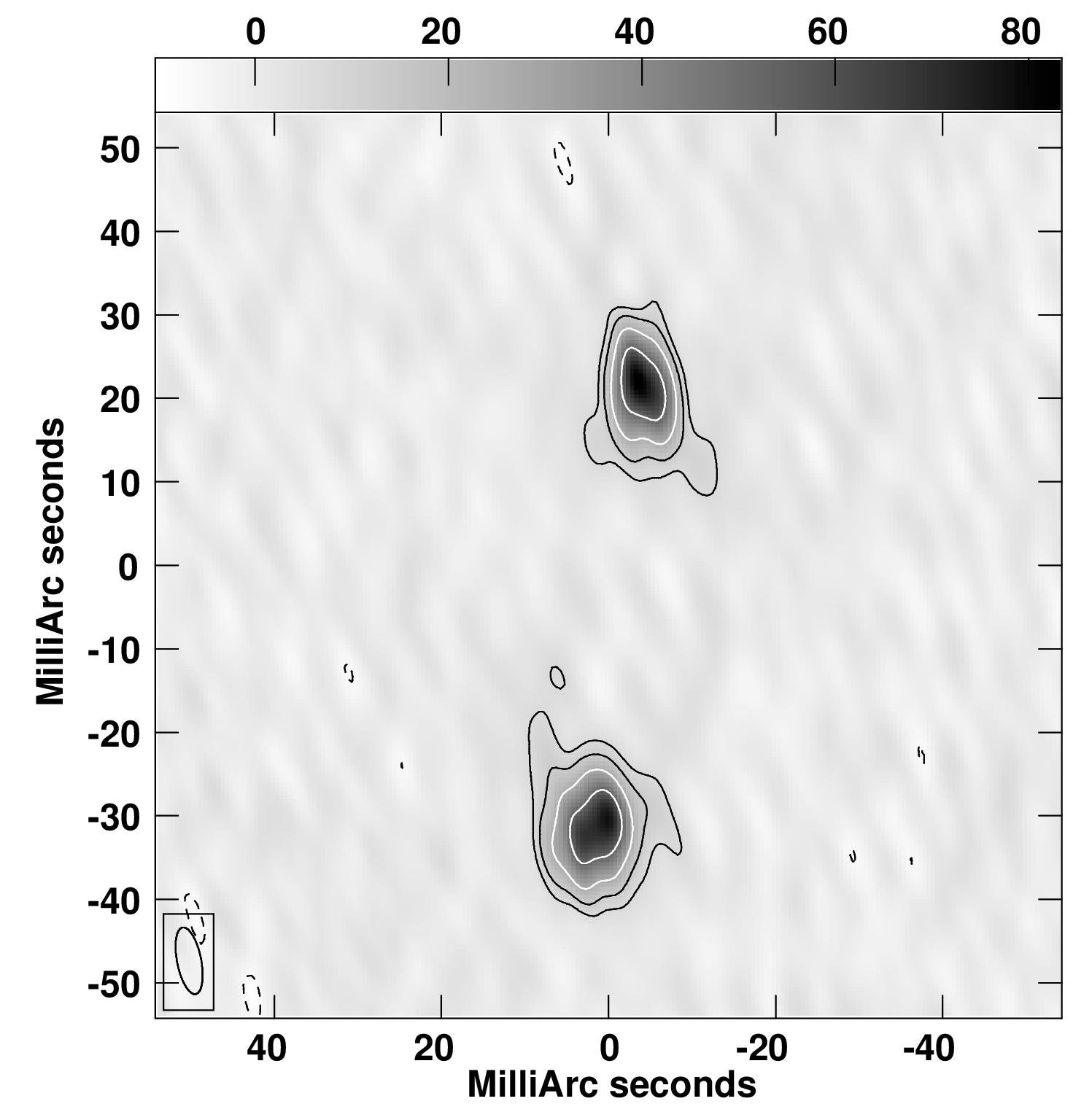}
    \includegraphics[scale=0.15]{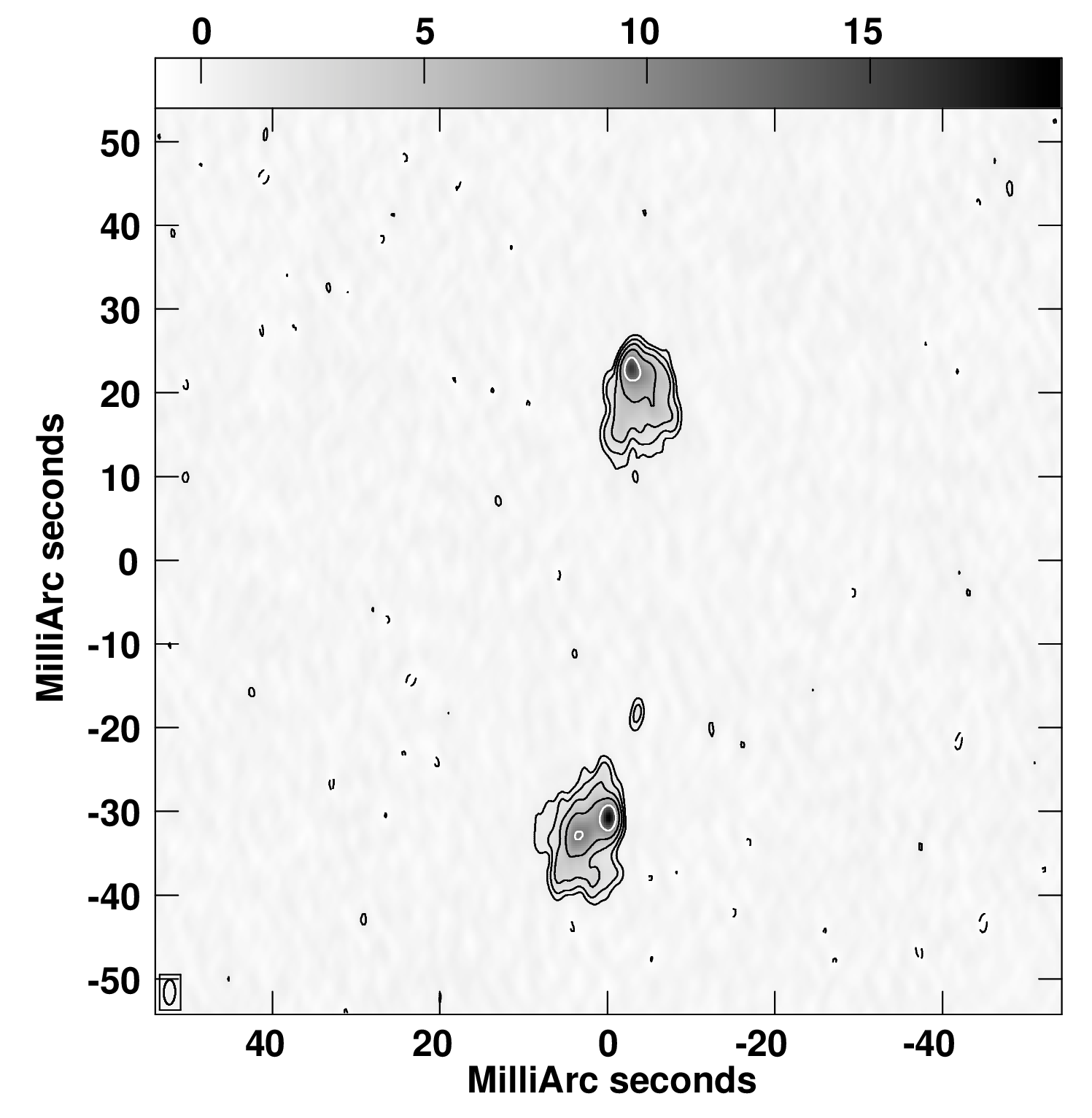}
    \caption{Images of J0855+5751 at 2.3 (left) and 5~GHz (right) made from archival VLBA data. The 2.3-GHz data are taken from the VLBA Calibrator Survey \citep{beasley02} and at 5~GHz from VIPS \citep{taylor05,helmboldt07}. Our new analysis of these data phase-referenced both images to the position of the 8-arcmin distant source J0854+5757. The restoring beams are shown in the bottom-left corner and have dimensions of $8.1 \times 2.9$~mas$^2$ at 11\fdg0 (2.3~GHz) and $3.0 \times 1.4$~mas$^2$ at $-$1\fdg4 (5~GHz). Both maps are centred at 08$^{\rmn{h}}$~55$^{\rmn{m}}$~21\fs357, $+$57$\degr$~51\arcmin~44\farcs10 (J2000). Contours are plotted at multiples ($-1$, 1, 2, 4, 8, 16, etc.) of 3$\sigma$ where $\sigma$ is the off-source rms noise in the map ($\sigma_{2.3} = 2.2$~mJy\,beam$^{-1}$, $\sigma_{5} = 230 \, \mu$Jy\,beam$^{-1}$). The greyscale shows source brightness in mJy\,beam$^{-1}$. The source is very likely a CSO \citep{taylor05} with the faint central component (visible at 5~GHz only) corresponding to the location of the radio core.}
    \label{fig:vlba}
  \end{center}
\end{figure*}

J0855+5751 is a similar system that has been discovered in the course of a Green Bank Telescope (GBT) survey for small impact parameter pairs of optical galaxies and background radio-loud quasars \citep{zwaan15}. The dwarf galaxy SDSS~J085519.05+575140.7 at a redshift of $z=0.026$ displays strong 21-cm absorption with a peak optical depth of 24~per~cent. The background source, J0855+5751, has a 1.4-GHz flux density of 636~mJy \citep{white97} and a projected distance from the foreground galaxy of 6.8~kpc, a typical impact parameter for \ion{H}{i} absorption detected in the gas disk of a galaxy with a luminosity of 0.05~$L^*$ \citep{zwaan05a}. VLBA observations at 2.3 \citep{beasley02} and 5~GHz \citep{helmboldt07} have demonstrated that this source consists of two resolved components separated by about 60~mas (Fig.~\ref{fig:vlba}). On the basis of the observed VLBI structure, \citet{taylor05} have classified this source as a Compact Symmetric Object (CSO).

Here we present global VLBI observations of J0855+5751 that have been used to map the \ion{H}{i} absorption in the foreground galaxy at high angular resolution, $\sim$4~mas, thus probing the spatial distribution of cold gas on scales ranging from 2 to 35~pc. In addition, we report on spectroscopy carried out with the William Herschel Telescope (WHT) with which we have measured the redshift of the background radio source's host galaxy.

Throughout this paper we assume a flat $\Lambda$CDM cosmology and perform calculations using Ned Wright's Javascript Cosmology Calculator \citep{wright06}.

\section{Observations and data reduction}
\label{observations}

\subsection{Global VLBI}

VLBI observations (project code GZ013) took place over a period of 13~h during 2013 June 05-06 using a global 17-station array. This comprised the 10 antennas of the VLBA, plus the European VLBI Network (EVN) antennas at Effelsberg, Jodrell Bank (Lovell), Westerbork (operating as a phased array), Torun, Medicina, Zelenchukskaya and Badary. Data were recorded in 4 dual circularly-polarized subbands, each with a width of 2~MHz. Three of these measured the continuum, whilst the remaining one was centred on the absorption line (1384.6~MHz).

We did not phase reference, partly because J0855+5751 is bright enough to self-calibrate, but also because a bright ($\sim$1~Jy) calibrator source, J0854+5757 (B0850+581), is located only 8~arcmin away. This therefore lay within the primary beam of all the antennas, apart from Westerbork which has a very small primary beam of only a few arcsec when functioning as a phased array. Nevertheless, short pointed observations of J0854+5757 were made at regular intervals. The bandpass calibrator, J0834+5534, was observed at multiple hour angles for a total of about one hour.

The data were correlated using the SFXC software correlator \citep{kettenis10} at the Joint Institute for VLBI in Europe (JIVE) with three separate passes. Two produced pseudo-continuum data sets with 16 channels per subband. One pass (``continuum'') correlated each source at its actual position, whilst the other (``in\_beam'') correlated each J0855+5751 scan at the position of the in-beam phase calibrator. The third pass (``line'') only included the subband containing the absorption and used 2048 channels. With Hanning smoothing, this is equivalent to a velocity resolution of $\simeq$420~m\,s$^{-1}$. A correlator averaging time of 4~s was chosen, this being preferred over 2~s in order to increase time smearing away from the phase centre and thus to reduce contamination by the in-beam calibrator.

Data reduction was performed using AIPS and proceeded initially using the ``in\_beam'' correlator data. These were subjected to standard VLBI reduction procedures, including {\em a priori} amplitude calibration using system temperatures and gain curves, antenna-based delay correction and bandpass calibration. For all antennas apart from Westerbork, the phases were then flattened over the entire time range in both time and frequency by fringe-fitting the data assuming a point-source model. The calibration tables were then copied to the ``continuum'' pass data so that J0855+5751 could be mapped and all antennas self-calibrated. As expected, the in-beam phase referencing produced an excellent initial map.

The final calibration was then transferred to the ``line'' dataset and a full-resolution antenna bandpass calculated. As Doppler tracking had not been used during the observations, the small time-variable changes in the line frequency as a result of the Earth's motion around the Sun were removed using the task {\sc cvel}. Finally, the continuum was removed by baseline fitting in the $(u,v)$ plane and a final map of the absorption in each channel made using the continuum-subtracted data. As the absorption line is quite narrow, the final spectral-line cube contains only 256 of the correlated spectral points.

\subsection{Archival VLBA data}

In order to properly compare our global VLBI data with previous continuum observations at other frequencies (2.3 and 5~GHz), we have retrieved the archival VLBA data\footnote{The project codes are BB023 \citep{beasley02} and BF072 \citep{helmboldt07} for the 2.3 and 5-GHz data respectively. We refer the reader to the original publications for details of the observations and data reduction.} and performed our own reduction. This followed standard procedures, but differed significantly from the original reductions in a number of ways. Firstly, although no phase-reference source was explicitly included at either frequency, the same calibrator used in the global VLBI observations (J0854+5757) had been included in each as a separate science target, in adjacent scans. Therefore we have been able to use this to phase reference the J0855+5751 data. Another difference concerns the 5-GHz data only. The published map \citep{helmboldt07} was produced using an automatic pipeline and contains obvious image defects such as large negative troughs around the lobes. Our new reduction was mapped and self-calibrated by hand and does not suffer from these issues.

\subsection{William Herschel Telescope}

J0855+5751 was observed on 20 April 2015 using the ISIS spectrograph on the 4.2-m William Herschel Telescope (service proposal SW2015a04). The red arm used the R158R grating for a total of three 15-min exposures, whilst in the blue arm a single 30-min exposure with the R300B grating was taken. Numerous bias frames and flat fields were taken, as well as exposures of simultaneous copper-argon and copper-neon arc lamps for wavelength-calibration purposes. The slit width was 1~arcsec. In what follows, only the data from the red arm are considered -- the source is very faint in the blue and these data have therefore not been analysed.

All data reduction was performed using standard methods with software provided as part of Starlink, particularly the Pamela package. Each exposure was initially debiased and flat-fielded, the galaxy's spectrum was traced on the chip, the sky emission estimated and the spectrum extracted using Marsh's optimal method \citep{marsh89} which includes rejection of pixels affected by cosmic rays. The three spectra were then combined and the wavelength scale calibrated. Flux calibration was not carried out as this is unnecessary for redshift determination.

\section{Results}

\subsection{Continuum images of the background source}

Our phase-referenced continuum images of the re-reduced 2.3 and 5~GHz data are shown in Fig.~\ref{fig:vlba}. The 2.3-GHz image shows the basic structure of the source, two prominent lobe-like components separated by $\sim$60~mas, but the 5-GHz image shows more detail in both. Of particular interest is the southern component which contains two distinct peaks and shows that faint, extended emission extends further south. There is also a very faint ($\sim 6\sigma$) compact source that lies between the two lobes, but closer to the southern component.

\begin{figure}
  \begin{center}
    \includegraphics[scale=0.15]{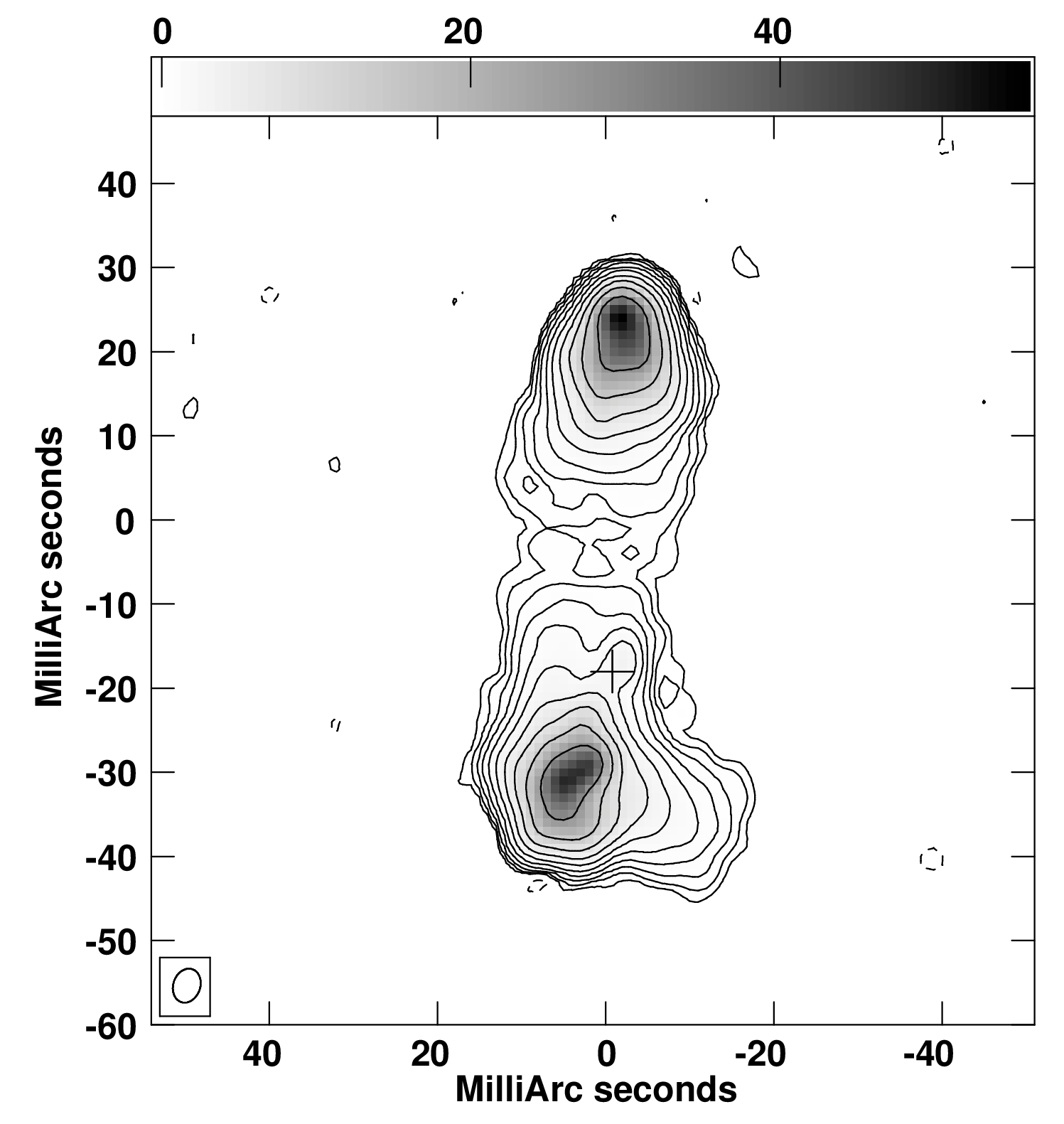}
    \caption{Global VLBI continuum map of J0855+5751 at an average frequency of 1386~MHz. The restoring beam has a size of $4.1 \times 3.2$~mas$^2$ at a position angle of $-$18\fdg8. The map phase centre and size is the same as that of the VLBA-only maps shown in Fig.~\ref{fig:vlba}, but offset slightly in order to centre the source better. Contours are plotted at multiples ($-1$, 1, 2, 4, 8, 16, etc.) of 3$\sigma$, where $\sigma$ is the off-source rms noise in the map (18~$\mu$Jy\,beam$^{-1}$).  The greyscale shows source brightness in mJy\,beam$^{-1}$. The faint feature just north of the southern lobe at 5~GHz is marked with a cross and lies close to a peak in the 1386-MHz map.}
    \label{fig:gz013_cont}
  \end{center}
\end{figure}

The new 1386-MHz continuum map of J0855+5751 is shown in Fig.~\ref{fig:gz013_cont} and achieves a 1-$\sigma$ rms noise of 18~$\mu$Jy\,beam$^{-1}$ and an angular resolution of $4.1 \times 3.2$~mas$^2$. The structure of the source at 1386~MHz is very similar to that seen at the higher frequencies, with the bulk of the emission contained in two lobe-like components. These seem to have similar shapes at all three frequencies, for example in the southern lobe, where the peak emission at 1386~MHz has a similar position angle ($-60$~degrees North through East) to that defined by the two peaks visible at 2.3 and 5~GHz.

However, the combination of the lower observing frequency and the high sensitivity in the new map reveal significantly more emission than was previously detected. This lies predominantly between the two components detected at 2.3~GHz, the brightness of the source at 1386~MHz declining relatively smoothly from peaks in each lobe (the maximum brightness is 56~mJy\,beam$^{-1}$ at the tip of the northern lobe) to much fainter emission in the middle of the source, where a bridge of material may connect the two lobes.

The faint central component between the two lobes seen at 5~GHz seems to correspond to a distinct peak in the 1386-MHz map and the 5-GHz position is marked with a cross in Fig.~\ref{fig:gz013_cont}. The observed offset between the peak position at each frequency (1.3~mas) may be due to errors in the phase-referencing or frequency-dependent structure, but these are certainly the same component. Measuring its spectral index\footnote{We use the following convention: $S_{\nu} \propto \nu^{\alpha}$.}, $\alpha$, is complicated by the surrounding lobe emission in the 1386-MHz map and the very different synthesized beams, but an upper limit can be measured using the peak brightness in each map. This is equal to 3.4 and 1.8~mJy\,beam$^{-1}$ at 1386~MHz and 5~GHz respectively and thus we estimate that $\alpha > -0.5$.

Another interesting feature of the new map is the spur of emission that protrudes from the western edge of the southern lobe. It points in a south-westerly direction, has a length of about 20~mas (as measured from the peak of the southern lobe) and is also resolved along its width.

\subsection{HI absorption in the foreground galaxy}

\begin{figure*}
  \begin{center}
    \includegraphics[scale=0.33]{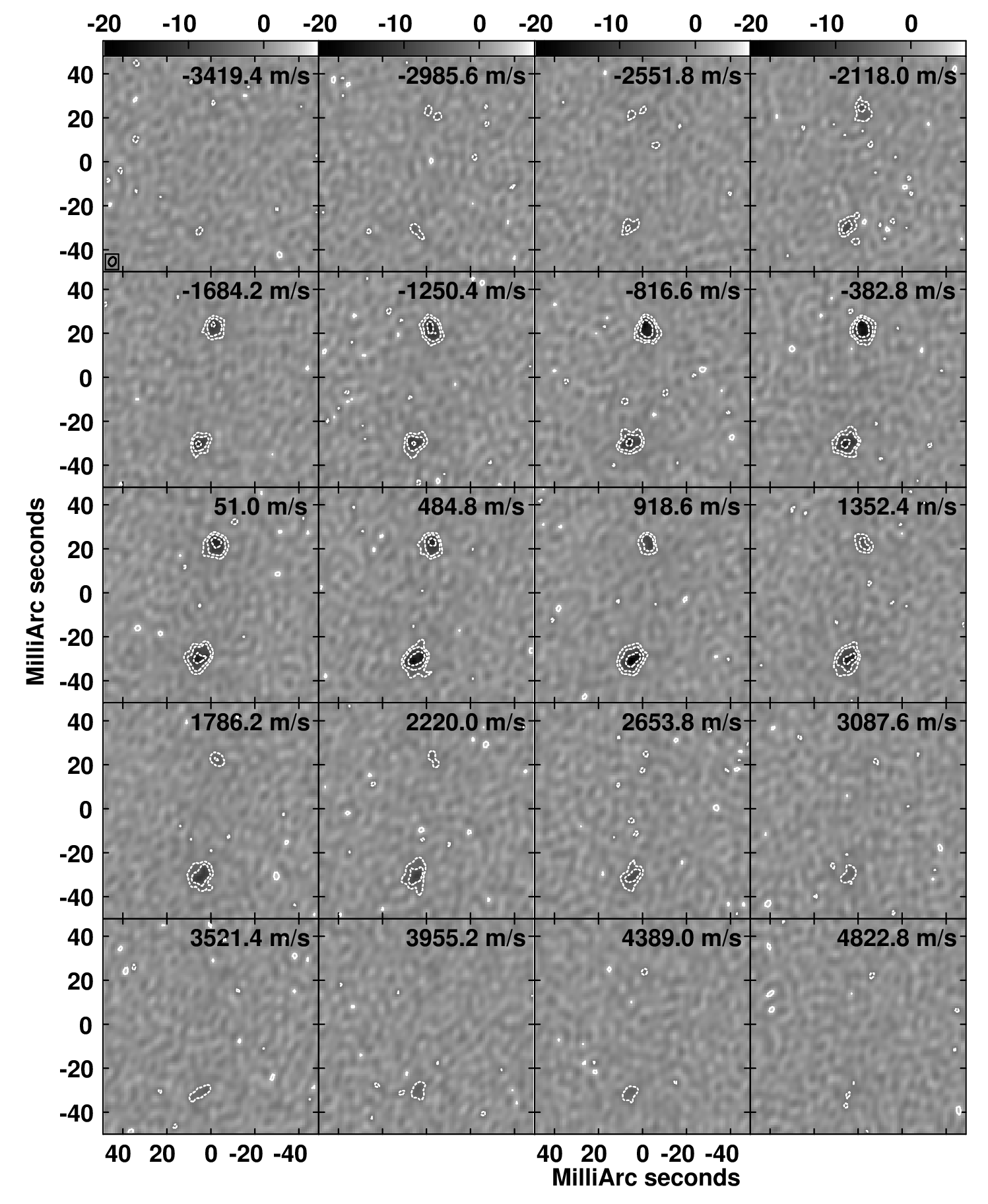}
    \caption{Maps of the continuum-subtracted data for 20 channels which cover the absorption profile -- only alternate channels are plotted and each is labelled with its heliocentric velocity (optical definition) relative to the peak in the GBT spectrum (7737.6\kms). Contours are plotted at $1, -1, -2, -4$ times 3$\sigma$, where $\sigma$ is the off-source rms noise in the map (1~mJy\,beam$^{-1}$). The greyscale is also in mJy\,beam$^{-1}$. Although the absorption becomes apparent in both lobes at approximately the same lower velocity, the absorption in the northern lobe becomes undetectable before that in its southern counterpart.}
    \label{fig:gz013_chmap}
  \end{center}
\end{figure*}

Channel maps of the continuum-subtracted data are shown in Fig.~\ref{fig:gz013_chmap}. The velocity range (8.25\kms) covers the full width of the absorption profile and only alternate channels are plotted for clarity. The labels in each panel show the heliocentric velocity (optical definition) of that channel and are given relative to 7737.6\kms.

\begin{figure*}
  \begin{center}
    \begin{minipage}[][][b]{0.7\columnwidth}
      \includegraphics[scale=0.3]{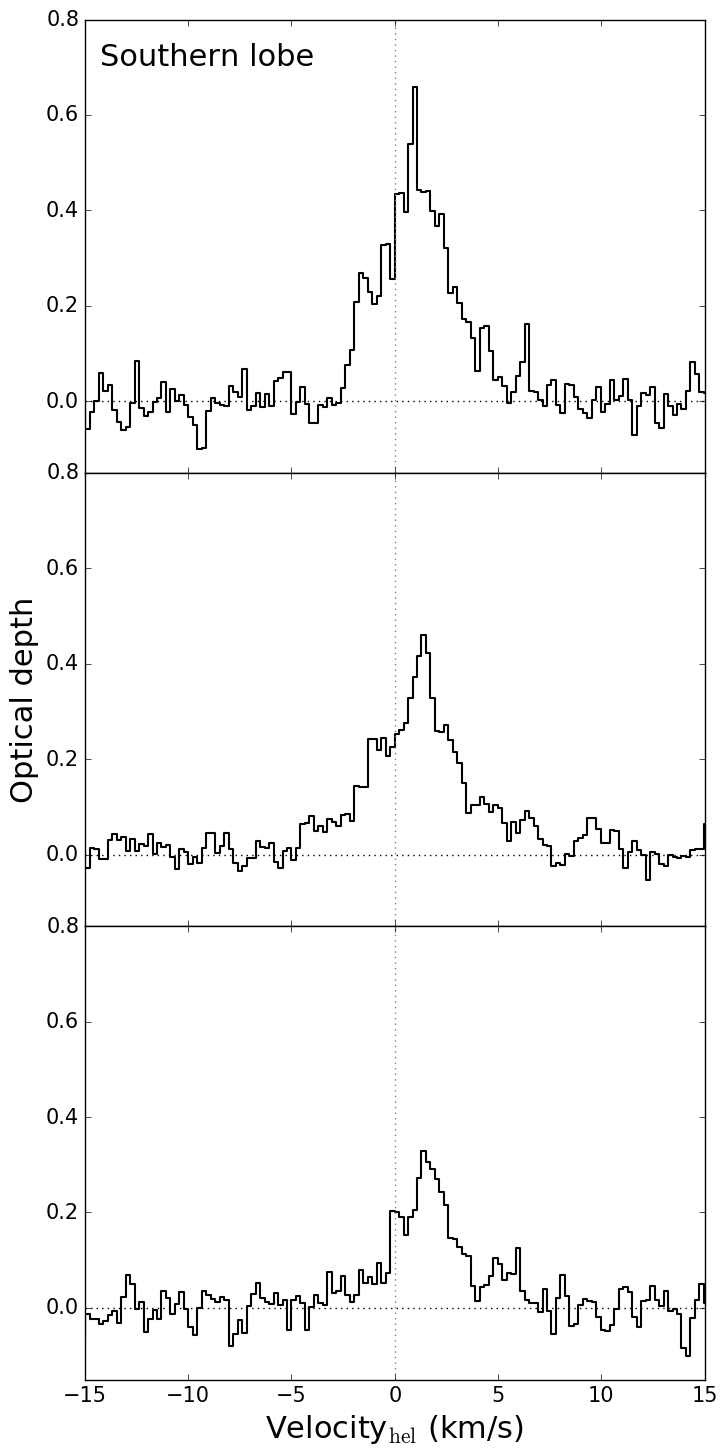}
    \end{minipage}
    \begin{minipage}{0.65\columnwidth}
      \includegraphics[scale=0.3]{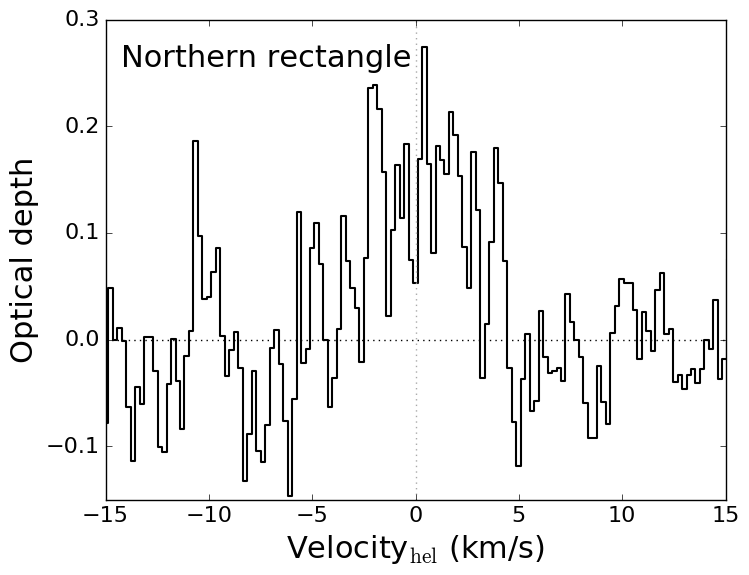}
      \includegraphics[scale=0.38]{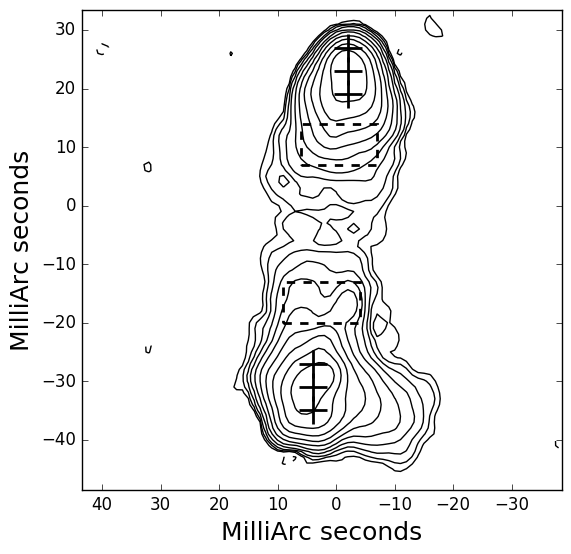}
      \includegraphics[scale=0.3]{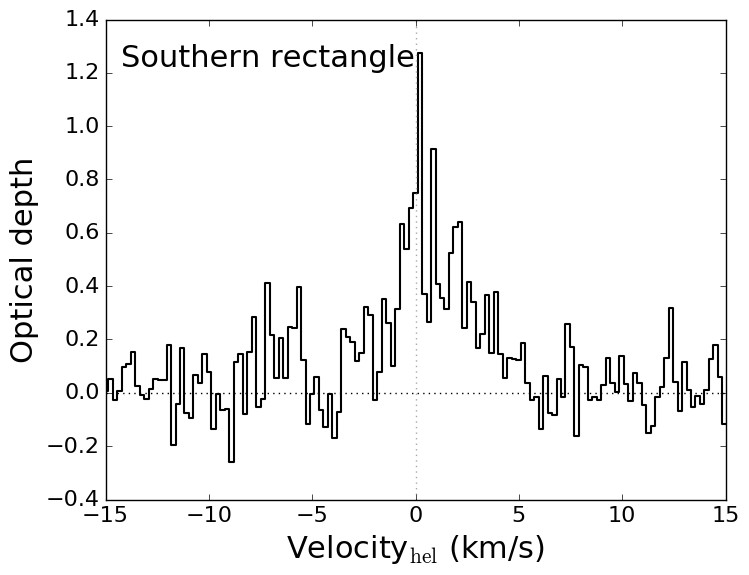}
    \end{minipage}
    \begin{minipage}[][][t]{0.7\columnwidth}
      \includegraphics[scale=0.3]{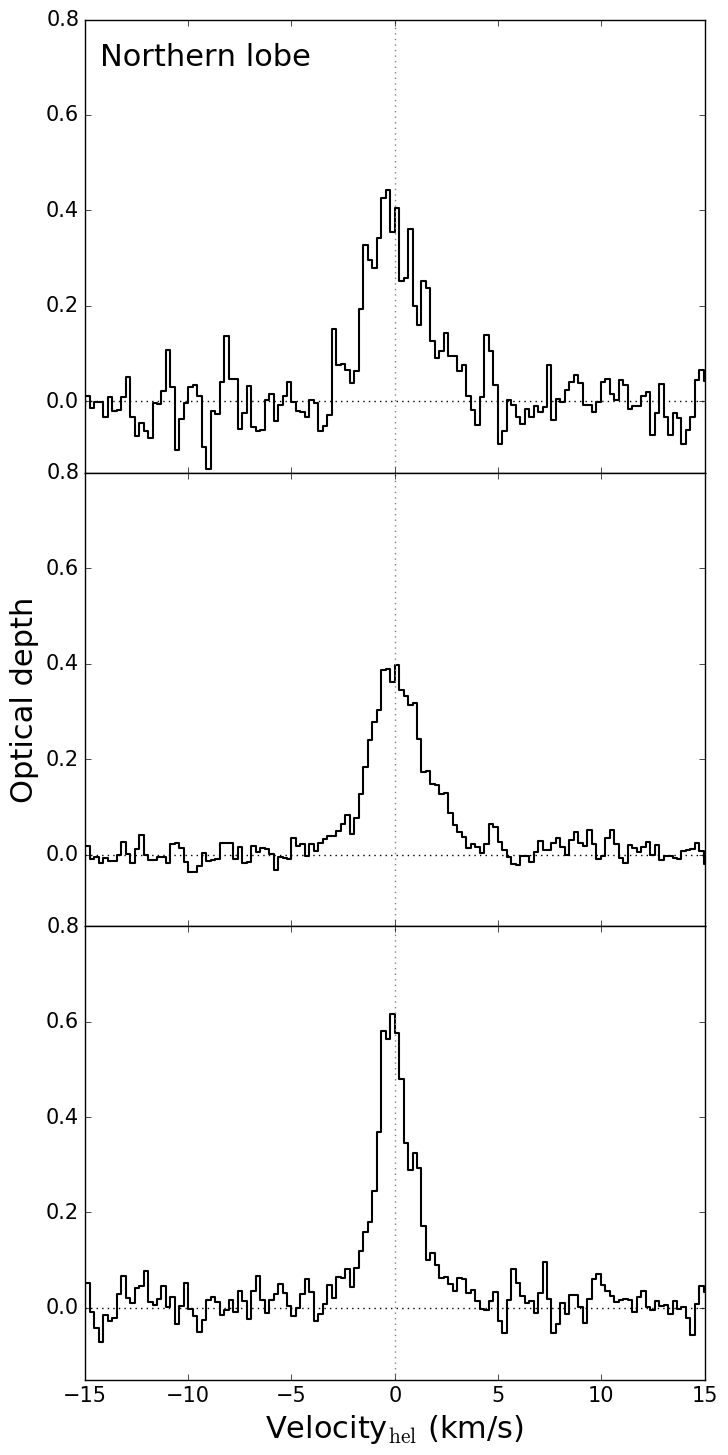}
    \end{minipage}
    \caption{Optical-depth spectra of three representative sightlines towards the northern (right) and southern (left) lobes. Their positions are marked on the continuum map and each spectrum is plotted on the same $y$-axis scale for easy comparison. The absorption profile varies significantly as a function of sight line to J0855+5751 and there is a significant velocity offset between the northern and southern components (all velocities are referenced to the peak of the GBT spectrum). Also shown are spectra derived from averaging the absorption within two rectangular areas within the fainter emission of each lobe. Although noisier, absorption is clearly detected in both and suggests that absorption is present across the entire source i.e.\ the covering factor is 1.}
\label{fig:gz013_tau}
\end{center}
\end{figure*}

Both lobes are clearly detected in absorption with a maximum depth of 21~mJy\,beam$^{-1}$ relative to the continuum. Each lobe is also resolved and has a shape similar to that seen in the continuum map, although the lower sensitivity per channel (1~mJy\,beam$^{-1}$) precludes a clear detection of absorption against the weaker continuum components. One notable feature of Fig.~\ref{fig:gz013_chmap} is that, although the absorption in both the northern and southern components extends to approximately the same negative velocity, towards positive velocities the northern component fades significantly before that of the southern i.e.\ the absorption profile seems to be narrower in the northern lobe.

Spectra of optical depth, $\tau$, for each spatial pixel in the map have been created using the standard equation,
\begin{equation}
  \label {eq:tau}
  \tau = - \ln \left(1 - \frac{\Delta S}{S} \right),
\end{equation}
where $S$ is the continuum flux density and $\Delta S$ the absorbed flux density. For very weak continuum components, random noise fluctuations can lead to the channel flux density being greater than the continuum and, as this would lead to logarithms of negative numbers, Equation~\ref{eq:tau} is modified (H. Liszt, private communication) to
\begin{equation}
  \label {eq:taumod}
  \tau = \ln \left(\left| 1 - \frac{\Delta S}{S} \right| \right).
\end{equation}
A selection of representative optical-depth spectra is shown in Fig.~\ref{fig:gz013_tau}, together with their position against J0855+5751. For each lobe, three sight lines along a north-south axis are displayed. These are separated from each other by the size of the synthesized beam and are therefore independent measures of the absorption profile.

Also shown are optical-depth spectra measured against two rectangular regions in the fainter emission between the brightest areas of each lobe. As the continuum here is much lower, single-pixel optical-depth spectra are very noisy and thus an average is necessary in order to gain a more reliable measurement. Due to the low signal-to-noise ratio (SNR) the flux-density absorption spectra and continuum are averaged separately before calculating a single optical-depth spectrum. Although noisier than the individual spectra in the brighter parts of the lobes, absorption is clearly detected.

\subsection{Redshift of the background source}

\begin{figure*}
  \begin{center}
    \includegraphics[scale=0.43]{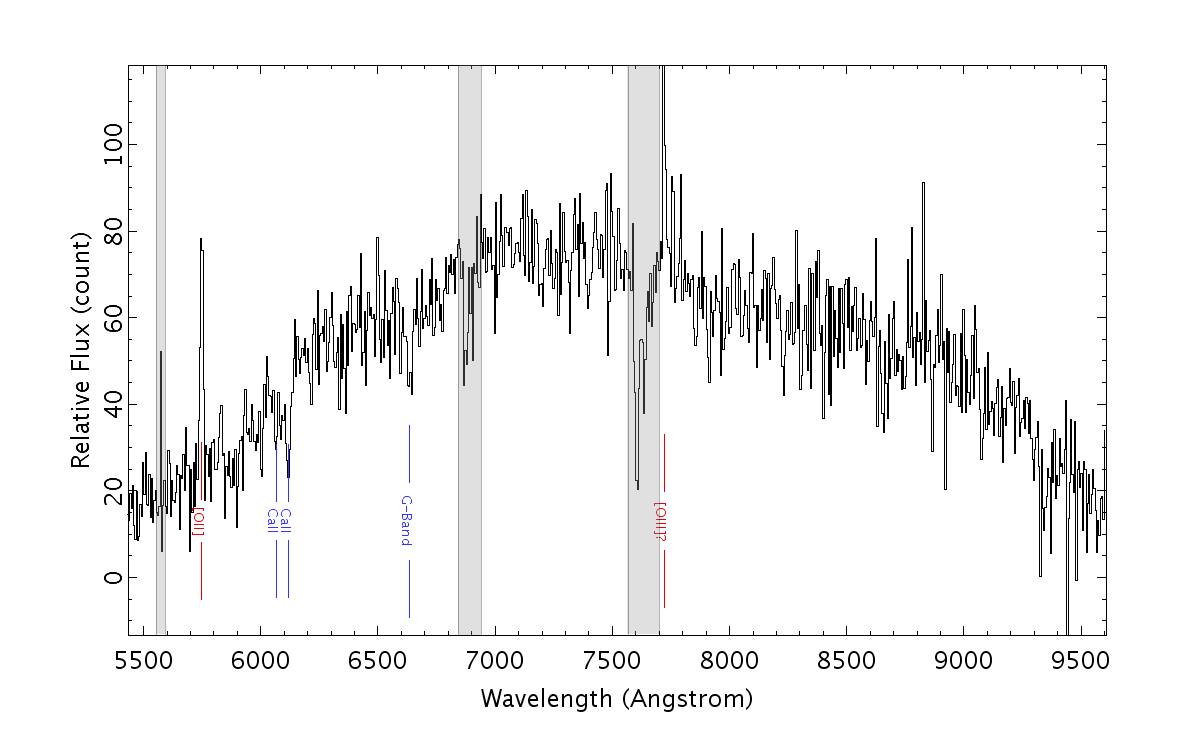}
    \caption{Optical spectrum taken with the red arm of ISIS on the William Herschel Telescope. Identifiable features in the spectrum include the [\ion{O}{ii}] emission line, as well as the \ion{Ca}{ii} H and K absorption lines and the G absorption band. These yield an unambiguous redshift of $z = 0.54186 \pm 0.00009$. The vertical grey bands mask out residual sky emission or absorption features. No flux calibration has been performed.}
    \label{fig:wht}
  \end{center}
\end{figure*}

The optical spectrum of J0855+5751 (re-binned by a factor of three) is shown in Fig.~\ref{fig:wht}. Although only one prominent emission line is clearly detected, its identity as the (blended) [\ion{O}{ii}] doublet at 3726.16 and 3728.91~\AA\ is in no doubt due to the presence of the \ion{Ca}{ii} H and K lines, as well as the G absorption band. It is possible that the [\ion{O}{iii}] line at 5007~\AA\ is also present, but this falls under a bright sky emission line and may not be real. The redshift was determined by fitting Gaussians to the [\ion{O}{ii}] line and absorption features using {\sc splat} and least-square fitting. We calculate a redshift for J0855+571 of $z = 0.54186 \pm 0.00009$.

\section{Discussion}
\label{discussion}

\subsection{The absorbing gas on 15 to 35-pc scales}

In contrast to the single-dish (GBT) spectrum \citep{zwaan15} which is dominated by a single Gaussian component, the optical-depth plots shown in Fig.~\ref{fig:gz013_tau} reveal a more diverse \ion{H}{i} velocity structure, particularly as seen against the southern lobe. This highlights the fact that any spectrum is always an average of the absorption contained within the angular extent of the telescope beam. However, it also emphasizes that VLBI, with its extremely high angular resolution, is less susceptible to this effect and can reveal a more complex and multi-component ISM than would otherwise have been assumed to be the case.

To illustrate this, for each lobe we can take the average of the spectra contained in multiple spatial pixels. As with the rectangular regions in Fig.~\ref{fig:gz013_tau}, we do this by first averaging the flux-density absorption spectra and continuum separately and then calculating the optical depth. In order to remove the faint emission between the lobes, we mask all pixels with a flux less than 15~$\sigma$.

\begin{figure*}
  \begin{center}
    \includegraphics[scale=0.43]{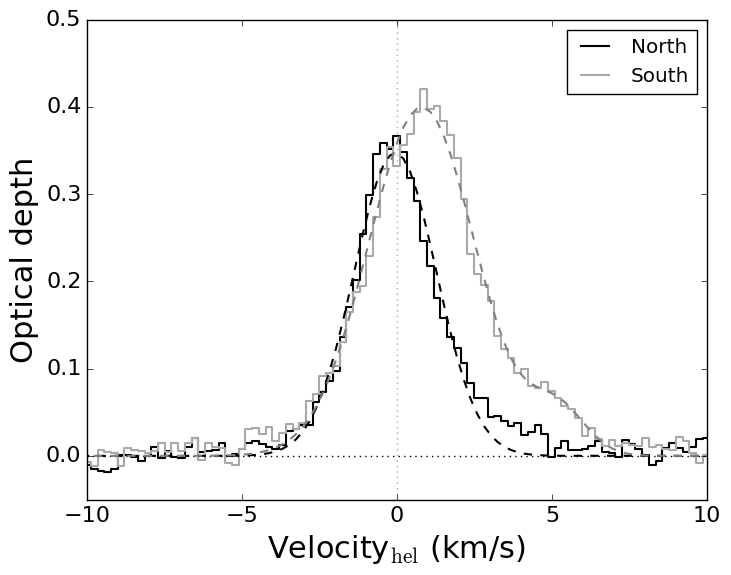}
    \includegraphics[scale=0.43]{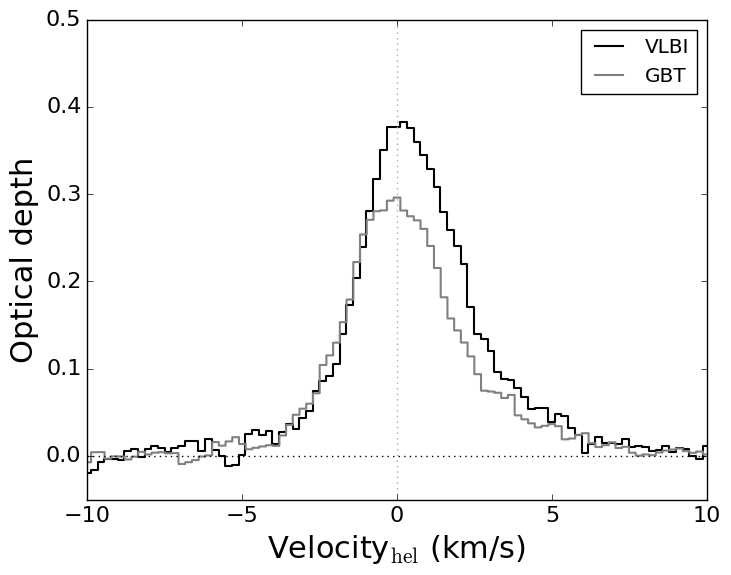}
    \caption{Left: Average optical-depth spectra derived from pixels with a continuum flux $>15~\sigma$ in the northern (black) and southern (grey) lobes. There is a clear velocity offset between the two and the gas seen against the southern lobe has a larger velocity dispersion and integral optical depth. Gaussian fits to each spectrum (Table~\ref{tab:lobe_gauss}) are shown as dashed lines. Right: VLBI spectrum averaged over both lobes (black) and the GBT spectrum (grey). The GBT optical depth has been corrected for the dilution caused by the presence of the VLBI phase calibrator in the GBT primary beam. All velocities are referenced to the peak of the GBT spectrum.}
    \label{fig:tau_lobes}
  \end{center}
\end{figure*}

\begin{table*}
  \caption{Parameters derived from Gaussian fitting to the summed pixels corresponding to each lobe (Fig.~\ref{fig:tau_lobes}). All errors correspond to the formal fitting uncertainties and do not include any contribution from systematic sources of error.}
  \label{tab:lobe_gauss}
  \begin{center}
    \begin{tabular}{cccccc} \\ \hline
      Component & $v$ (\kms) & $\sigma_v$ (\kms) & $T_{\text{k}}$ (K) & $\int \tau dv$ (\kms) & $N_{\text{HI}}$ ($\times 10^{20} \text{cm}^{-2}$) \\ \hline
      North  & $-0.05\pm0.02$ & $1.33\pm0.02$ & $213.9\pm2.5$ & $1.16\pm0.02$ & $<$4.5 \\
      South1 & $0.80\pm0.02$ & $1.72\pm0.03$ & $358.9\pm4.5$ & $1.72\pm0.03$ & $<$11.3 \\
      South2 & $5.17\pm0.13$ & $0.91\pm0.13$ & $99.9\pm11.8$ & $0.12\pm0.02$ & $<$0.2 \\ \hline
    \end{tabular}
  \end{center}
\end{table*}

The average optical-depth spectra are shown in Fig.~\ref{fig:tau_lobes}. As expected, both are smoother than the individual pixel spectra shown in Fig.~\ref{fig:gz013_tau} and look more like the single-dish spectrum i.e.\ each is approximately Gaussian and exhibits excess signal at higher recession velocities. Despite the broad similarities, the two are clearly different, particularly in terms of the relative narrowness of the northern line which was already evident in the individual frequency maps in Fig.~\ref{fig:gz013_chmap}. We have parameterized each by fitting Gaussians, two to the southern line profile (South1 and South2) but only one to the northern counterpart due to the lower SNR in the high-velocity gas. The fits are shown together with the spectra in Fig.~\ref{fig:tau_lobes} and the fit parameters are given in Table~\ref{tab:lobe_gauss}. 

Assuming that the line is purely thermally broadened, it is possible to use the measured velocity dispersions to calculate upper limits on the kinetic temperature of the \ion{H}{i} gas using
\begin{equation}
  \label{eq:kint}
  T_{\text{k}} \le \frac{m_{\text{H}} \Delta v^2}{k \, 8 \ln 2} = \frac{1.2119 \times 10^2 \Delta v^2}{8 \ln 2}
\end{equation}
where $k$ is the Boltzmann constant, $m_{\text{H}}$ the mass of the hydrogen atom and $\Delta v$ the full width at half maximum (FWHM) of the Gaussian fit. In turn, a measurement of $T_{\text{k}}$ allows the column density of neutral Hydrogen, $N_\mathrm{HI}$, to be determined where
\begin{equation}
  \label{eq:hicol}
  N_{\text{HI}} = 1.823 \times 10^{18} \, (T_s/f) \int \tau dv
\end{equation}
and $T_s$ represents the electron spin temperature (assumed to be equal to $T_{\text{k}}$ for collisionally-excited gas) and $f$ the covering factor (assumed to be unity). The values of $T_{\text{k}}$ and $N_{\text{HI}}$ are also given in Table~\ref{tab:lobe_gauss}.

\citet{zwaan15} measure a velocity dispersion of 1.5\kms\ from their single-dish spectrum, but our VLBI data show that the \ion{H}{i} absorbing gas seen towards the northern lobe is narrower than this ($\sigma_v = 1.33$\kms). This allows us to determine a tighter upper limit on the kinetic gas temperature of $T_{\text{k}} <$214~K, compared to the single-dish value of $T_{\text{k}} <$275~K. For the weaker South2 component, the upper limit is lower still, $T_{\text{k}} <$100~K. These temperatures also leave no doubt that we are observing the CNM.

Being able to separate the two lobes shows that there is more absorbing gas in front of the southern lobe, where the total optical depth is 50~per~cent higher than in front of the northern lobe. In terms of \ion{H}{i} column density, the upper limit for the South1 component is nearly six times higher than the threshold, $N_{\text{HI}} > 2 \times 10^{20}$~cm$^{-2}$, usually taken as indicating a Damped Ly-$\alpha$ absorber \citep[DLA;][]{wolfe05}, and is significantly in excess of the single-dish value.

We have also summed the absorption and continuum across the entire radio source by including all pixels with a continuum flux $>5 \sigma$. This can be compared with the GBT spectrum, but it is necessary to first correct this for the presence of the VLBI phase calibrator in the GBT primary beam. This will have increased the continuum level observed by the GBT and thus diluted the apparent optical depth. Although variable, two decades of monitoring by the University of Michigan Radio Observatory demonstrate that this does not exceed 10~per~cent at their lowest frequency of 5~GHz. The NRAO VLA Sky Survey \citep{condon98} 1.4-GHz flux density of 1.1~Jy is therefore probably quite reliable and, together with a GBT primary-beam attenuation of 85~per~cent\footnote{Determined using equations in the GBT Proposer's Guide.}, implies an optical-depth correction factor of 1.25.

The VLBI and corrected GBT spectra are shown in Fig.~\ref{fig:tau_lobes} where it can be seen that although the line shapes are broadly similar, the VLBI spectrum generally has a higher optical depth, particularly at higher velocities. That the optical depths at low velocities agree quite well suggests that the dilution correction has been successful and therefore that the VLBI maps are missing faint and/or extended emission that has relatively low optical depth. That flux has been missed is supported by the fact that the total flux density of the VLBI map is $\la$90~per~cent of the expected, although we caution that the flux density scale of VLBI data is rarely accurate to better than 5~per~cent.

\subsection{The absorbing gas on 2 to 10-pc scales}

\begin{figure*}
  \begin{center}
    \includegraphics[scale=0.49]{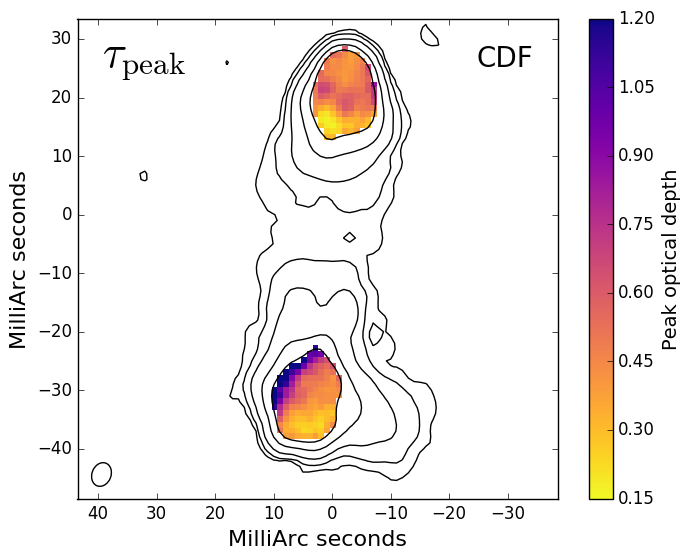}
    \includegraphics[scale=0.49]{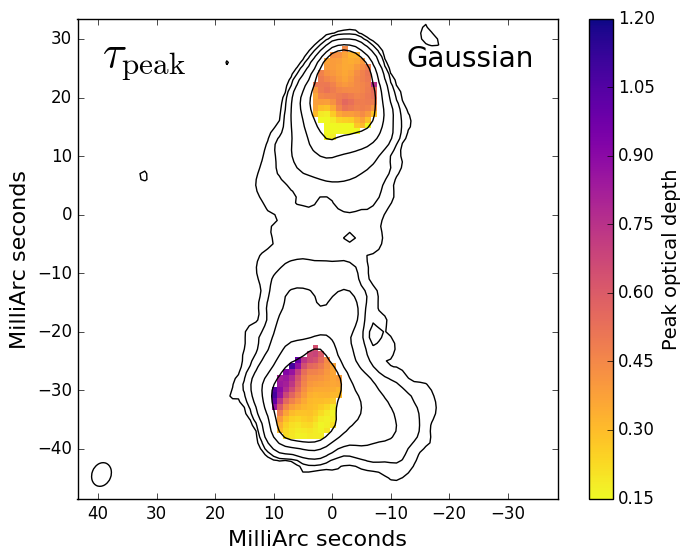}
    \includegraphics[scale=0.49]{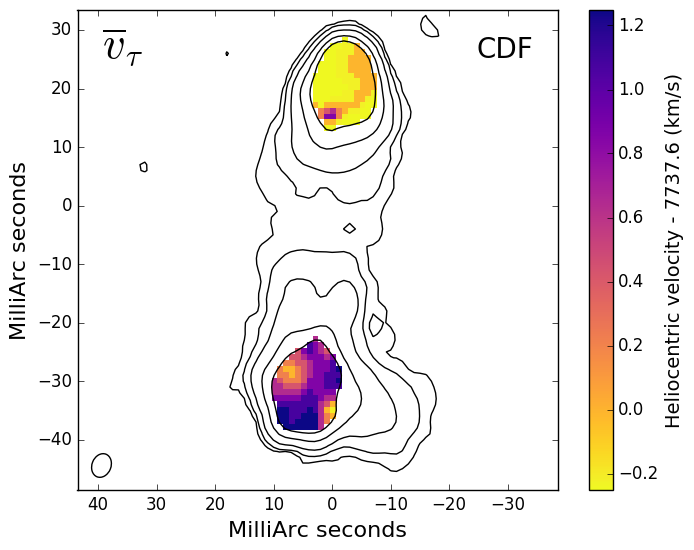}
    \includegraphics[scale=0.49]{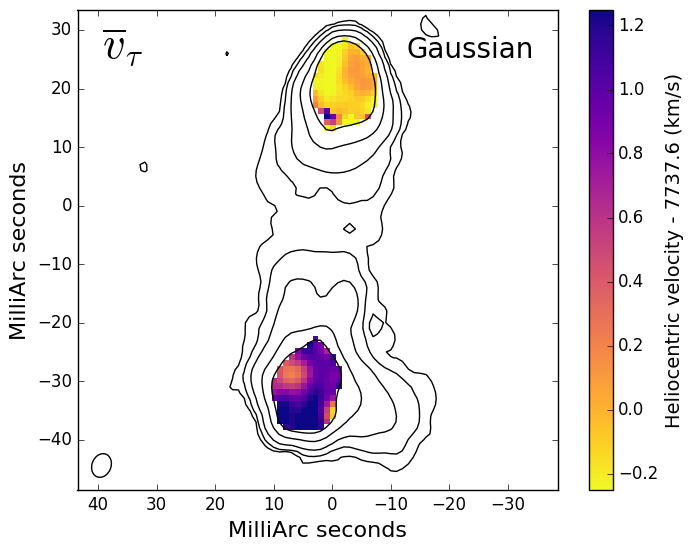}
    \includegraphics[scale=0.49]{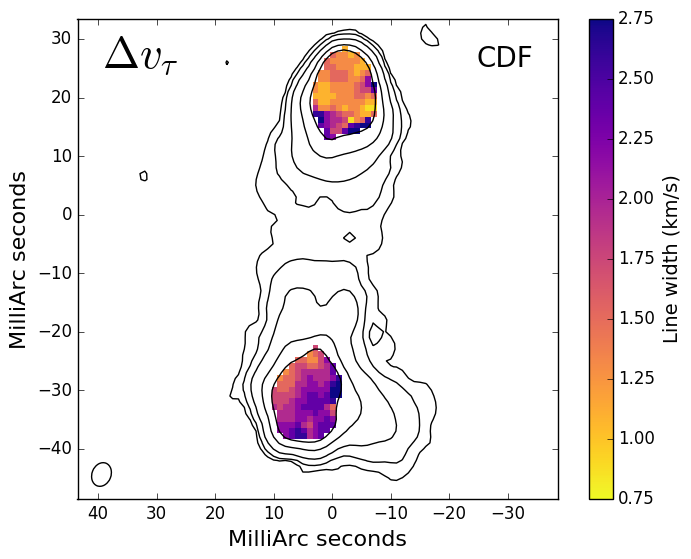}
    \includegraphics[scale=0.49]{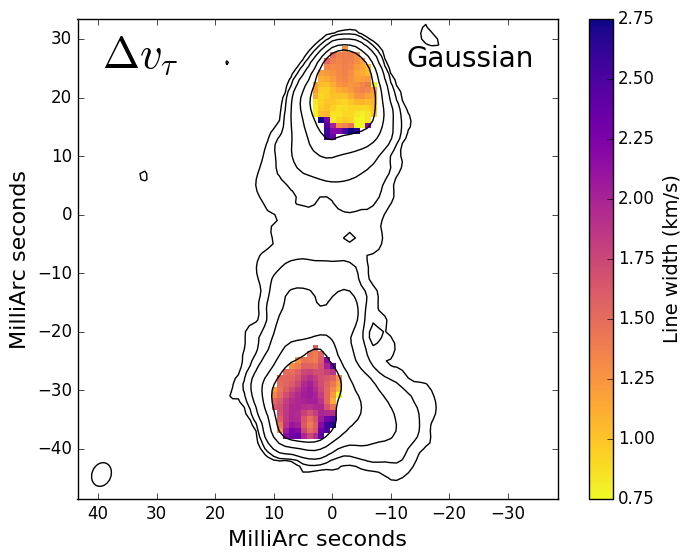}
    \caption{Absorption-line characteristics as a function of spatial coordinate: peak optical depth (top), velocity of maximum optical depth (middle) and width of absorption-line profile (bottom). The left column shows the values for the CDF method and the right column for the Gaussian fitting. The line width of the CDF method is equivalent to the $\sigma$ of the Gaussian fit. Also plotted are several contours of the continuum map in order to guide the eye -- the highest contour corresponds to the continuum threshold ($\ge9$~mJy\,beam$^{-1}$) used to include pixels. The restoring beam is shown in the bottom-left corner of each plot. Each pair of plots uses the same $z$-axis range for ease of comparison and it can be seen that both the CDF and Gaussian line-parameterization methods give similar results. The values at the southern edge of the northern lobe are not reliable as the optical depth here is very low.}
\label{fig:linemap}
\end{center}
\end{figure*}

\begin{figure*}
  \begin{center}
    \includegraphics[scale=0.49]{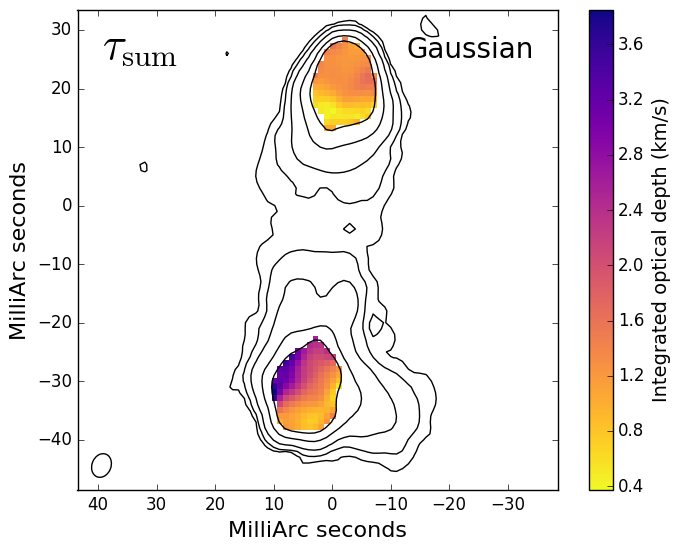}
    \includegraphics[scale=0.49]{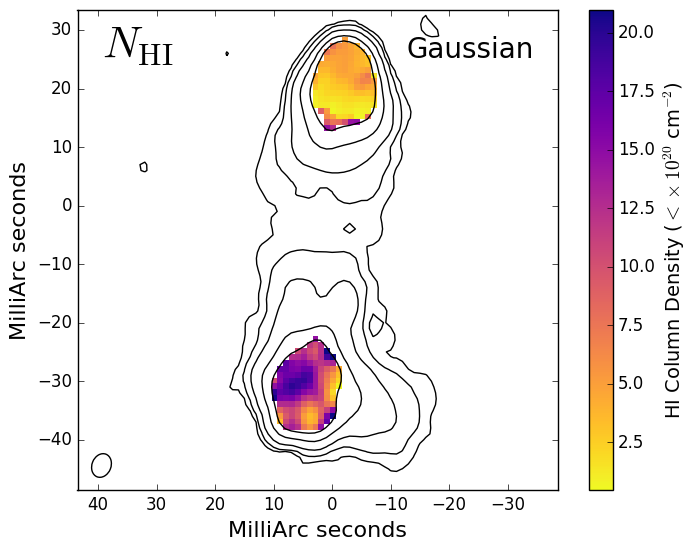}
    \caption{Plots of total optical depth (left) and \ion{H}{i} column density (right) as a function of spatial coordinate, for the Gaussian method only. The $N_{\text{HI}}$ values at the southern edge of the northern lobe are not reliable as the optical depth here is very low and the line width is poorly constrained. The threshold for considering an object a DLA is $N_{\text{HI}} > 2 \times 10^{20}$~cm$^{-2}$.}
\label{fig:optdmap}
\end{center}
\end{figure*}

Analysing the \ion{H}{i} properties on the smallest linear scales (2~pc) made possible by the very high angular resolution of our VLBI observations can be achieved by parameterizing the absorption profiles corresponding to each spatial pixel and plotting these parameters as two-dimensional maps. This can be done in various ways, for example by fitting Gaussians to each pixel's spectrum \citep*[e.g.][]{peck99}. Although the spectra shown in Fig.~\ref{fig:gz013_tau} are only approximately Gaussian, this technique offers the benefit that a small number of parameters are optimized and the fits are relatively robust against the presence of noisy data. However, as the spectra are not perfectly described by Gaussians, the use of non-parameterized methods can potentially give more accurate results for e.g.\ the total optical depth in each spectrum.

With this in mind, we have devised a technique whereby the line parameters are estimated using the cumulative distribution function (CDF) of the optical depth as a function of frequency. This essentially sums pixels across the line location (as determined from the single-dish spectrum) and thus measures the total optical depth, $\tau_{\text{sum}}$. However, it also allows a determination of equivalents to all the parameters measured by a Gaussian fit. The line width ($\Delta v_{\tau}$) is calculated as half of the difference between the velocities encompassing the central 68.3~per~cent of the optical depth. The average velocity ($\overline{v}_{\tau}$) is that where the CDF reaches 50~per~cent of its maximum value and the peak optical depth ($\tau_{\text{peak}}$) is simply the location of the peak pixel in the spectrum.

Ultimately, we use both estimators as a check on our results and show maps of the parameters that measure the peak, width and position of the line in Fig.~\ref{fig:linemap}. In order to only include those pixels where the line parameters are relatively well constrained, a continuum threshold of $\ge9$~mJy\,beam$^{-1}$ is imposed, which results in a mapped area equivalent to $\sim$26 VLBI synthesized beams. Over this region the results from the two methods are qualitatively very similar, each showing similar spatial variations. The spectra are also similar quantitatively, as can be seen from the use of the same $z$-axis range for each method. One obvious difference is in $\tau_{\text{peak}}$, where the lower values found using the Gaussian method demonstrate that noise spikes are indeed less influencing the result. In fact, the greater smoothness of the Gaussian maps in general is probably a consequence of the parameterized fits being more robust to the presence of noise.

As the two methods yield such consistent results, we therefore have a slight preference for the results of the Gaussian fitting and accordingly show in Fig.~\ref{fig:optdmap} the total optical depth ($\tau_{\text{sum}}$) and upper limit to the total \ion{H}{i} column density ($N_{\text{HI}}$) for this estimator alone. However, we emphasize that the quantitative results are also not that different and in fact the two $\tau_{\text{sum}}$ maps are almost identical.

\begin{figure*}
  \begin{center}
    \includegraphics[scale=0.6]{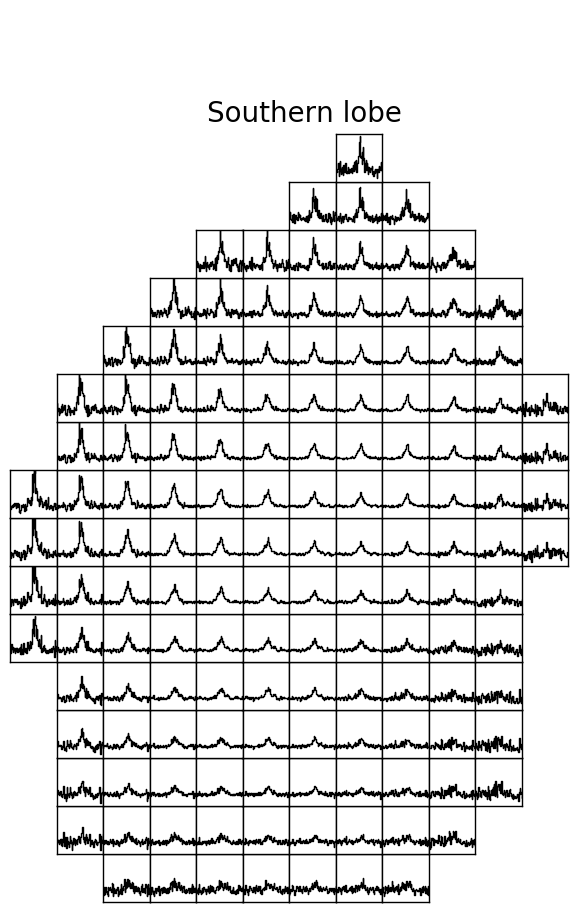}
    \includegraphics[scale=0.6]{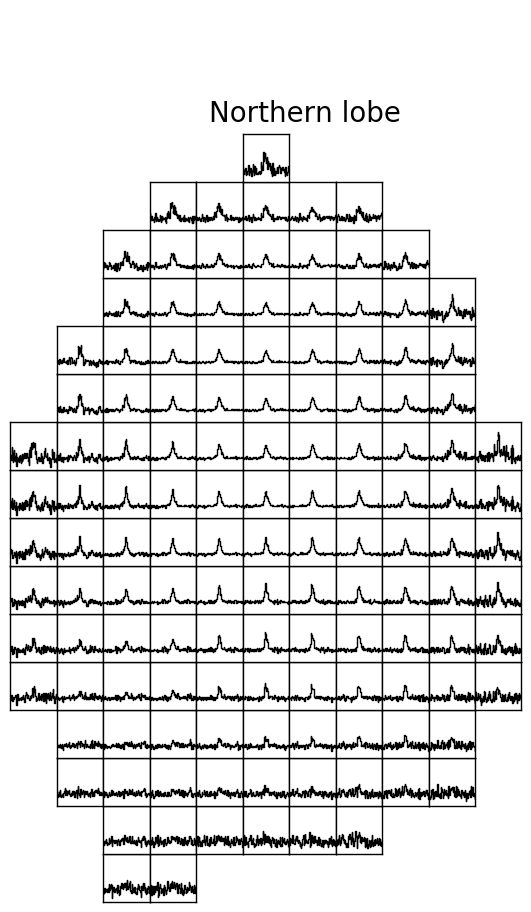}
    \caption{Individual optical-depth spectra for the southern (left) and northern (right) lobes. Each spectrum corresponds to a pixel displayed in Figs~\ref{fig:linemap} and \ref{fig:optdmap} and each is separated by 1~mas from its neighbour in both the vertical and horizontal direction. In the northern lobe the line gradually weakens towards the South-West, whilst in the southern lobe it strengthens towards the North-East. The x-axis range is $-15$ to 15\kms and optical depths are plotted between -0.4 and 1.2.}
\label{fig:all_spectra}
\end{center}
\end{figure*}

Figs~\ref{fig:linemap} and \ref{fig:optdmap} both show differences between the lobes that were apparent from the average of each lobe (Fig.~\ref{fig:tau_lobes}) but variations across each lobe are also now readily apparent. The gas seen against the northern lobe is relatively uniform, although the optical depth does decrease significantly at the southern boundary. The reduction in absorption-line strength makes it more difficult to reliably determine the line properties and this is probably the reason for the sudden jump in line width and velocity at this location. Another view of the fading of the absorption is provided by Fig.~\ref{fig:all_spectra} where the spectrum at each pixel is plotted as a function of its location in the map.

In contrast, the gas seen against the southern lobe shows significant variations across its entire extent. The absorption velocity is relatively constant across most of the lobe, but a region to the North-East with a size of a few synthesized beams is offset by approximately $-1$\kms. Large and apparently coherent variations in line width are visible and the total optical depth increases by a factor of $\sim$5 between the westerly and easterly edges. This is equivalent to an angle of about 12~mas, or a linear distance at $z=0.026$ of 6.3~pc. As expected, the large parsec-scale fluctuations revealed by our optical-depth maps and spectra translate into similarly large variations in the distribution of \ion{H}{i} column density (Fig.~\ref{fig:optdmap}).

The observed large increase is not an artefact of our analysis, nor due to higher noise at lower continuum fluxes. Both the Gaussian and CDF methods give the same result and the Gaussian fitting in particular is less affected by noise spikes in the spectra. The increase in absorption towards the East is particularly obvious in Fig.~\ref{fig:all_spectra} where it can also be seen that the line strength does not correlate with noise -- our maps correctly identify areas of high or low optical depth.

We therefore see structure on all spatial scales probed by our data. The \ion{H}{i} properties are very different between the gas seen against the two lobes (i.e.\ on scales up to $\sim$30~pc) whilst there is also significant structure on scales of order the size of our synthesized beam (2~pc). Additionally, we find no evidence for sightlines that do not intercept absorbing \ion{H}{i} gas. Although our maps and spectra show that the line becomes very weak at the southern end of the northern lobe, the averaged spectrum from the rectangular area further south in this lobe (Fig.~\ref{fig:gz013_tau}) shows that low optical-depth absorption remains. Similarly, a clear detection of absorption is also made in the northern part of the southern lobe.

It seems therefore that the CSO J0855+5751 is seen against a single coherent cold structure, which must therefore have an extent along one axis of at least 35~pc. As this is considerably larger than the largest CNM clouds featured in the \citet{mckee77} model, this provides further evidence that the blobby sheet model of \citet{heiles03}, as well as being a better description of the ISM of our own galaxy, is also appropriate for external galaxies. Although the properties of the absorption lines can be satisfactorily modelled with Gaussians, it is clear from the high SNR spectra that there are deviations from non-Gaussianity, which suggests that we are actually seeing an average of multiple components. We interpret peaks seen within the broader line profile as individual cold cores (blobs) located within the larger-scale sheet.

Our observation of significant variation of the \ion{H}{i} optical depth on parsec scales in an external galaxy has implications for measuring the spin temperature of absorbing gas in DLAs. The Ly-$\alpha$ line gives a direct measurement of the \ion{H}{i} column density which, in conjunction with the optical depth measured from the 21-cm line, enables a measurement of $T_s$. For this to be valid, one must be sure that the gas probed by the optical and radio observations is the same. \citet{kanekar14} use smoothed \ion{H}{i} emission maps of the Large Magellanic Cloud (LMC) to demonstrate that, for low-redshift ($z<2$) DLAs, $N_{\text{HI}}$ is relatively constant over the size of the radio source and thus that estimates of $T_s$ are probably within 10~per~cent of the true value. However, the angular resolution of the unsmoothed LMC map was only 15~pc, whilst our analysis emphasizes that the \ion{H}{i} can be unevenly distributed on smaller scales. Given the exceptionally small size of the continuum-emitting region of optical quasars, we therefore caution that $T_s$ values obtained using this method may in fact be significantly less reliable than suggested.

\subsection{A jet-cloud collision in the background CSO}

Based on the 5-GHz image, \citet{taylor05} identified J0855+5751 as a CSO and our work strengthens this conclusion. The definition of a CSO includes a linear size of less than 1~kpc \citep{wilkinson94} and, with the redshift now known, we can confirm that for J0855+5751 this is equal to 400~pc. The lobe-dominated symmetry of CSOs is thought to be a consequence of the source lying in the plane of the sky and thus a weak core component lying between the two lobes is usually expected. We identify this as the faint feature located between the two lobes at 5~GHz. If this is the core, it should have a flat or inverted spectrum and indeed its spectral index ($\alpha > -0.5$) is significantly flatter than that of the source as a whole ($\alpha \sim -0.8$).

If this component is the core and the source indeed lies close to the plane of the sky, it is perhaps surprising that the core is located much closer to the southern lobe and not midway between the two. A possible explanation for this lies in the complicated structure of the southern lobe. The northern lobe looks like an archetypal radio lobe, with a well-defined hotspot furthest from the core and more extended emission which declines in surface brightness towards the core. This is a classic edge-brightened FR~II morphology \citep{fanaroff74} and we note that this is consistent with the source's total radio luminosity. Using the 151-MHz flux density \citep{hales07} and the source redshift, we measure $L_{\text{151~MHz}} = 1.2 \times 10^{27}$~W\,m$^{-2}$ which is well above the $\sim 2 \times 10^{25}$~W\,m$^{-2}$ threshold that marks the approximate division between the two classes.

In contrast, the 5-GHz image shows very clearly that the southern lobe contains multiple peaks, the brightest of which lies closer to the core. There are hints of more extended emission to both the north and south of these and, in the new 1386-MHz map, so much more extended emission is visible that the southern emission resembles a classical FR~II edge-brightened radio lobe. However, the prominent spur of emission that protrudes from the western edge again demonstrates the significantly distorted structure compared to the northern lobe.

One explanation for this is that the southern jet encounters some obstacle, perhaps a dense gas cloud, in the interstellar medium of the host galaxy. This has prevented the southern hotspot (the brightest and most compact feature) from moving any further from the core (or significantly reduced its velocity) whilst the northern hotspot has continued its advance. The extended emission to the north of the southern hotspot is therefore typical lobe emission that is usually seen behind hotspots in FR~II sources. We identify the south-westerly protrusion as jet emission that has been deflected from its original path.

Distorted lobes of the type observed in J0855+5751 are certainly not unique to this source, similar examples including 3C~254 \citep{thomasson06} and 4C~29.50 \citep{lonsdale86}. We note that in both cases the presumed location of the jet-cloud interaction also lies much closer to the radio core than the emission constituting the other lobe. In the case of 3C~254, there is additionally spectroscopic evidence of an interaction between the jet and gas in the galaxy \citep{bremer97}. \citet{odea98} lists a number of compact radio sources which display distorted morphology, including some examples where there is evidence other from just the radio morphology.

\subsection{Absorption prospects in the background CSO}

CSOs are known to be gas-rich systems. The detection rate of associated \ion{H}{i} absorption systems is relatively high \citep{pihlstroem03,gupta06,chandola11} and the lack of radio polarization is presumably due to depolarization of the intrinsically-polarized emission as it passes through a magnetized plasma in the host galaxy. In addition, the detection of [\ion{O}{ii}] emission is a direct indicator of star formation in the host galaxy of J0855+5751 and the deflected jet in the southern lobe also strongly suggests the presence of a dense cloud of gas.

\begin{figure}
  \begin{center}
    \includegraphics[scale=0.44]{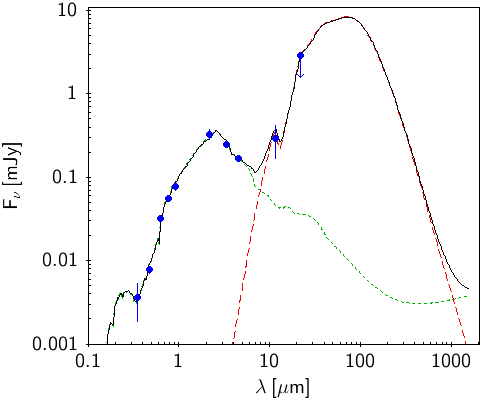}
    \caption{Optical/infrared spectral energy distribution of the host galaxy of J0855+5751 as a function of observed wavelength. The data are taken from the SDSS ($u$, $g$, $r$, $i$, $z$), 2MASS ($K$-band) and WISE (3.4, 4.6, 12 and 22\micron). The short- and long-dashed lines show the stellar and AGN SED fits respectively. Note that the WISE4 point is an upper limit.}
    \label{fig:0855_sed}
  \end{center}
\end{figure}

In order to investigate the global properties of the host galaxy in more detail, we have assembled its broad-band spectral energy distribution (SED) using SDSS Data Release 10 (DR10; \citealt{ahn14}) 2MASS \citep{skrutskie06} and WISE All-Sky data release photometry. Although J0855+5751 was not in the 2MASS catalogues, it was clearly visible in the $K_{\text{s}}$-band image and we extracted its flux manually using SExtractor. We then compared the observed SED to a series of models comprising two different components: a stellar component composed of simple stellar population (SSP) models of solar metallicity built using the Padova evolutionary tracks \citep{bertelli94} and a grid of AGN torus models described in \citet{nenkova08}. For a full description of the SED fitting and individual model components see \citet{hatzimi08,hatzimi09,feltre12}. We did not use a starburst template, as we have virtually no constraints on the far-infrared (FIR) part of the SED. The SED and fit are shown in Fig.~\ref{fig:0855_sed}.

The combined SDSS, 2MASS and WISE1 and WISE2 data points provide very strong constraints on the stellar component. The derived stellar mass of the galaxy is 4.8~$\times 10^{11}$\,M$_{\odot}$, with an estimated total extinction of the stellar population of E(B-V) = 0.19 and an age of 7.9~Gyr, corresponding to the oldest stellar population. Although the WISE3 data point is indicative of hot dust emission, the torus model is very poorly constrained. The mid-to-FIR part of the observed SED can be reproduced solely by a compact torus with a large covering factor and a high equatorial optical depth at 9.7\micron\ ($\tau_{9.7}=3.8$) although a starburst component cannot be ruled out. 

The SED is therefore consistent with the host galaxy of a powerful radio source that is heavily obscured, which is in turn consistent with CSOs being radio sources where the jets lie in the plane of the sky. We thus believe that, with the optical redshift now measured, J0855+5751 would make a promising target for further radio spectroscopy. This might also allow the detection of outflows, partly because of the observed jet-cloud interaction, but also because these are commonly observed in compact radio sources \citep*{morganti05,holt08}. However, a particularly exciting prospect is that, as well as searching for associated \ion{H}{i} absorption, molecular absorption (OH) might also be detectable.

Only five high-redshift molecular absorption systems have been identified to date, despite many attempts to increase this number \citep[e.g.][]{zwaan15,curran11,curran06}. From a study of the known OH systems, \citet{curran06} claim that the low number of current detections stems from relying too heavily on optical selections and that very red sources ($V - K \gtrsim 5$) should be selected. As well as being a radio-selected object, the optical SED can be seen to be very red and indeed the fitted SED gives $V - K = 5.1$. J0855+5751 thus appears to be a very good candidate for an OH absorption search.

\section{Conclusions}
\label{conclusions}

We have presented global VLBI observations of the bright radio source J0855+5751 which have allowed us to probe the properties of the \ion{H}{i} gas in the foreground dwarf galaxy SDSS~J085519.05+575140.7 at $z = 0.026$. We detect 21-cm absorption on all sightlines towards the background source where the SNR is high enough and, with a maximum velocity shift of $\le$2\kms, we seem to be probing a single coherent CNM structure with a minimum extent along one axis of 35~pc. The large size of this structure provides support for the applicability of the \citet{heiles03} ``blobby sheet'' model to the ISM of external galaxies.

Our very high angular resolution ($3-4$~mas) has allowed us to probe and map the variation in the 21-cm optical depth against the two lobes of the background radio source on linear scales as small as $2$~pc. Whilst the absorption properties of the northern gas are rather uniform, those of the southern gas are much less so, with the total optical depth varying by a factor of five across a distance of $\sim$6~pc. The upper limit on the \ion{H}{i} column density lies significantly above the threshold that defines a DLA. Our observation of significant optical-depth variations on scales of order 2~pc suggests that $T_s$ measurements from low-redshift ($z < 2$) DLAs may be significantly more unreliable than reported by \citet{kanekar14}.

The continuum map of the background radio source is by far the most sensitive yet made of J0855+5751 and detects much more of the extended emission in the radio lobes. The map includes a faint component previously seen at 5~GHz which, by virtue of its location and relatively flat spectrum, is probably the radio core. The new map thus strengthens the previous identification of this source as a Compact Symmetric Object. The southern lobe is distorted in comparison to its northern counterpart, most probably due to a collision with a dense cloud in the host galaxy. Given the CSO identification, future VLBI observations at 5~GHz might detect expansion of the source and lead to a measurement of a kinematic age.

The redshift of the host galaxy has been measured using optical spectroscopy and is equal to $z = 0.54186 \pm 0.00009$. This offers the possibility of searching for absorption associated with the radio source. Associated \ion{H}{i} absorption is often seen in CSOs and other compact radio sources and, with the redshift of the host galaxy now known, a search for this could be made in J0855+5751. In addition, the host galaxy's very red colour ($V - K \gtrsim 5$) raises the possibility of detecting OH absorption in this system. Unfortunately, RFI and a lack of VLBI arrays with suitable receivers make such searches at redshifts greater than $\sim$0.1 difficult. This situation will change with the advent of the Square Kilometre Array (SKA) as this will be based in very benign RFI environments, be capable of tuning to effectively any plausible redshifted frequency of \ion{H}{i} or OH and also have baselines up to 3000~km \citep{morganti15}.

\section*{Acknowledgements}

The European VLBI Network is a joint facility of European, Chinese, South African and other radio astronomy institutes funded by their national research councils. The National Radio Astronomy Observatory is a facility of the National Science Foundation operated under cooperative agreement by Associated Universities, Inc. The William Herschel Telescope and its service programme are operated on the island of La Palma by the Isaac Newton Group in the Spanish Observatorio del Roque de los Muchachos of the Instituto de Astrof\'{i}sica de Canarias. AIPS is produced and maintained by the National Radio Astronomy Observatory, a facility of the National Science Foundation operated under cooperative agreement by Associated Universities, Inc. This publication makes use of data products from the Wide-field Infrared Survey Explorer, which is a joint project of the University of California, Los Angeles, and the Jet Propulsion Laboratory/California Institute of Technology, funded by the National Aeronautics and Space Administration. This publication also makes use of data products from the Two Micron All Sky Survey, which is a joint project of the University of Massachusetts and the Infrared Processing and Analysis Center/California Institute of Technology, funded by the National Aeronautics and Space Administration and the National Science Foundation. The research leading to these results has received funding from the European Commission Seventh Framework Programme (FP/2007-2013) under grant agreement No. 283393 (RadioNet3). This research has made use of data from the University of Michigan Radio Astronomy Observatory which has been supported by the University of Michigan and by a series of grants from the National Science Foundation, most recently AST-0607523.

On a more personal level, we thank the staff of JIVE and the various observatories for their efforts, as well as Ian Browne for his help with the WHT proposal and Tom Marsh for his advice during the reduction of the WHT data. We also acknowledge the use of various scripts created by John Taylor that made the optical data analysis much easier. CP thanks the ESO science visitor program for support. When it comes to Python, Adam Ginsburg is a marvel. We are grateful to the anonymous referee whose comments and suggestions greatly improved the manuscript.




\bibliographystyle{mnras}
\bibliography{mybib}


\bsp	
\label{lastpage}
\end{document}